# The Eccentric Kozai-Lidov Effect as a Resonance Phenomenon

**Vladislav V. Sidorenko**

**Abstract** Exploring weakly perturbed Keplerian motion within the restricted three-body problem, Lidov (1962) and, independently, Kozai (1962) discovered coupled oscillations of eccentricity and inclination (the KL-cycles). Their classical studies were based on an integrable model of the secular evolution, obtained by double averaging of the disturbing function approximated with its first non-trivial term. This was the quadrupole term in the series expansion with respect to the ratio of the semimajor axis of the disturbed body to that of the disturbing body. If the next (octupole) term is kept in the expression for the disturbing function, long-term modulation of the KL-cycles can established (Ford et al. 2000; Naoz et al. 2011; Katz et al. 2011). Specifically, flips between the prograde and retrograde orbits become possible. Since such flips are observed only when the perturber has a non-zero eccentricity, the term "Eccentric Kozai-Lidov Effect" (or EKL-effect) was proposed by Lithwick and Naoz (2011) to specify such behaviour. We demonstrate that the EKL-effect can be interpreted as a resonance phenomenon. To this end, we write down the equations of motion in terms of "action-angle" variables emerging in the integrable Kozai-Lidov model. It turns out that for some initial values the resonance is degenerate and the usual "pendulum" approximation is insufficient to describe the evolution of the resonance phase. Analysis of the related bifurcations allows us to estimate the typical time between the successive flips for different parts of the phase space.

V.V.Sidorenko
Keldysh Institute of Applied Mathematics
Russian Academy of Sciences,
Miusskaya Sq., 4, 125047 Moscow, RUSSIA

Moscow Institute of Physics and Technology
Institutskiy S-Str., 9, 141700 Dolgoprudny, RUSSIA
E-mail: vvsidorenko@list.ru

*Dedicated to the loving
memory of Lyudmila*

**1 Introduction**

Consider the restricted three-body problem "star + test particle + planet", with both the planet and the particle orbiting the star. Under the assumption that the semimajor axis of the particle's osculating orbit, $a$, is much smaller than the semimajor axis of the planet's orbit, $a_p$, the disturbing function characterising the gravitational influence of the planet on the particle's motion can be expanded over powers of the ratio $a/a_p$ — and can then be approximated by a partial sum of that series.

Retaining only the first non-trivial term (i.e., the quadrupole term) in this expansion and averaging this term twice (over the orbital motion of the planet and of the test particle), we obtain an integrable model describing the secular evolution of the test particle's osculating orbit. This model was pioneered by Lidov (1962) and Kozai (1962). Exploring this model, these authors discovered independently that the eccentricity, $e$, and the inclination, $i$, of the particle's orbit exhibit long-period interrelated oscillations (the Kozai-Lidov cycles or, abbreviated: the KL cycles). These cycles are classified into rotating and librating types, dependent on the behaviour of the argument of the pericentre, $\omega$. The librating KL cycles sometimes appear in the literature under the name of the "Kozai-Lidov resonance".

The importance of the Kozai-Lidov effect follows from its universality. Various celestial bodies provide examples of this effect in their orbital motion: asteroids (Froeschle et al. 1991; Vashkovyak 1986), satellites of the giant planets (Nesvorny et al. 2003; Vashkovyak 1999, 2003), triple stars (Eggleton and Kiseleva-Eggleton 2001; Harrington 1968), and even black holes (Blaes et al. 2002; Wen 2003). A special role is given to the Kozai-Lidov effect in various scenarios describing the formation of the dynamical architecture of exoplanetary systems (Innanen et al. 1997; Libert and Henrard 2007). More examples of the Kozai-Lidov effect can be found in the recent book by Shevchenko (2016).

It should be noted that the Kozai-Lidov effect can be observed even in systems that fail to meet the assumptions imposed by Lidov (1962) and Kozai (1962). In (Bailey et al. 1992; Gronchi and Milani 1999; Kozai 1979; Thomas and Morbidelli 1996; Vashkovyak 1981), this effect was discovered in the dynamics of small bodies with semimajor axes comparable to the semimajor axis of the planets disturbing these bodies. It turned out that in the case of mean motion resonance the same phenomenon occurs (Kozai 1985). The Kozai-Lidov effect is observed also when the orbital motion of a celestial body is disturbed by a disk or some other extended astrophysical object (Ivanov et al. 2005; Subr and Karas 2005).

Ford et al. (2000), Naoz et al. (2011), and Katz et al. (2011) took a further step by keeping also the next, octupole, term from the expansion of disturbing function over the powers of $a/a_p$. In the case of the perturber's motion in orbit with the eccentricity $e_p > 0$, such a modification yields a modulation of the KL cycles; i.e., long-term variations in the maximal and minimal values of the eccentricities and inclinations over these cycles. Most notably, the inclusion of the octupole term renders the possibility of flips between the prograde ($i < 90°$) and retrograde ($i < 90°$) motion, as was discovered by (Naoz et al. 2011; Katz et al. 2011). Since such flips are observed only for $e_p > 0$, Lithwick and Naoz (2011) proposed to call this phenomenon the "eccentric Kozai-Lidov effect" (the EKL effect). Various aspects of the EKL effect are discussed in (Naoz 2016). In particular, it is expected that this effect can (under some additional hypotheses) explain the existence of the strange objects in the exoplanetary systems, which move opposite to their host star's spin (Triaud et al. 2010).

The aim of our paper is to demonstrate that the EKL effect can be interpreted in a meaningful way as a resonance phenomenon (Arnold et al. 2006). To this end, we rewrite in terms of "action-angle" variables the integrable model of weakly perturbed Keplerian motion proposed by Lidov and Kozai (Section 2). These variables enables us to further simplify the problem through one more averaging (that over the KL cycles). This is carried out in Section 3. Interpretation of the obtained results is given in Sections 4 and 5. In Appendix A we present some auxiliary formulae characterizing the Laplace vector behavior in rotating KL cycles. In Appendix B the conditions of the motion without flips are discussed.

*Remark 1.* In (Li et al. 2014b), the so called "low-inclination" EKL effect was studied. It is an extreme case of the EKL effect when the particle moves most of the time (except the instants of flips) in an orbit with a small inclination or in an orbit with the inclination close to $180°$. Technically, the "resonance" interpretation of the "low-inclination" EKL effect can be achieved in a simpler manner: one can use as a departure point the integrable dynamical model based on the double averaged equations of motion of a planar restricted elliptic three-body problem (Aksenov 1979). We consider it as possible direction for the future activity.

*Remark 2.* For an "outer" variant of the restricted elliptic three-body problem (i.e., in the case of $a/a_p \gg 1$), the possibility of periodic changes in the direction of the test particle's orbital motion has long been known (Farago and Laskar 2010; Ziglin 1975). It is noteworthy that in the "outer" problem the eccentricity of the particle's osculating orbit does not vary substantially, while in the "inner" problem the EKL effect furnishes singular orbits (at a certain instant of time, the osculating orbit turns into a segment of a straight line, with $e \approx 1$).

## 2 "Action-angle" variables for the Kozai-Lidov Hamiltonian

2.1 The Kozai-Lidov Hamiltonian

Let $K$ denote the Hamiltonian of the integrable 2DOF Hamiltonian system, introduced by Lidov (1962) and Kozai (1962) as a model of the Keplerian motion perturbed by a distant celestial body. To reveal more delicate dynamical phenomena (in particular, the EKL effect), one needs to employ a more refined model with the Hamiltonian $\mathcal{K}$ having the form

$$\mathcal{K} = K + \varepsilon K^*, \qquad (1)$$

where

$$\varepsilon = \frac{a}{a_p} \frac{e_p}{1 - e_p^2} \ll 1.$$

The second term on the right-hand side of (1) characterizes the "averaged" influence on the test particle's motion, produced by the octupole term in the series expansion of the disturbing function. Discussion of various aspects of the averaging procedure leading to model (1) is presented in (Luo et al. 2016; Naoz et al. 2013)

Under a proper choice of units, $K$ and $K^*$ can be written as

$$K(G, H, g, -) = -\frac{3}{8}\left\{H^2 + 2(1-G^2)\left[1 - \frac{5}{2}\left(1 - \frac{H^2}{G^2}\right)\sin^2 g\right]\right\}. \quad (2)$$

$$K^*(G, H, g, h) = \frac{75}{64}\left\{\frac{2H}{G}\sqrt{1 - \frac{H^2}{G^2}}\sin h - \sqrt{1-G^2}\left(\cos h \cos g - \sin g \sin h \cdot \frac{H}{G}\right)\right.$$
$$\left. \cdot \left(\frac{1}{5} - (1-G^2)\left[\frac{8}{5} - 7\sin^2 g\left(1 - \frac{H^2}{G^2}\right)\right] - H^2\right)\right\}.$$

Here $G = \sqrt{1-e^2}$ is a generalized momentum conjugated to the coordinate $g = \omega$, while $H = \cos i\sqrt{1-e^2}$ is a generalized momentum conjugated to the cyclic coordinate $h = \Omega$. As ever, $e$, $i$, $\omega$, and $\Omega$ are the osculating Keplerian elements of the test particle.

From the symmetry inherent in the system under consideration, it follows that

$$\mathcal{K}(G, -H, g, 2\pi - h) = \mathcal{K}(G, -H, 2\pi - g, h) = \mathcal{K}(G, H, g, h). \quad (3)$$

Keeping in mind further application of the standard perturbation technique developed to study resonance phenomena in near-integrable Hamiltonian systems (Arnold et al. 2006), we specify some properties of the unperturbed system ($\varepsilon = 0$), which will be important later.

To illustrate the secular effects, which can be described by the Kozai-Lidov model, we present in Figure 1 phase portraits of the 1DOF Hamiltonian system

$$\frac{dg}{dt} = \frac{\partial K}{\partial G}, \quad \frac{dG}{dt} = -\frac{\partial K}{\partial g}, \quad (4)$$

depending on $H$ as a parameter. In the case of $H \neq 0$, a typical trajectory on the phase portrait corresponds either to a librating or a rotating KL cycle. At $H = 0$, the solutions of the system (4) are of a singular nature: at a certain instant of time, the osculating orbit of the particle turns into a segment of a straight line (with $G \to 0$ and, respectively, $e \to 1$). Nevertheless, we can associate these regimes with the librating and rotating families of solutions as their formal limits for $H \to 0$.

Since the Hamiltonian $K$ is autonomous, its value $\varkappa$ remains constant along the trajectories serving as solutions to the equations of motion. To classify the qualitative properties of the secular evolution for different values of $H$ and $\varkappa$, we introduce the auxiliary functions

$$\varkappa_-(H) = \frac{3}{8}(H^2 - 2), \qquad \varkappa_0(H) = -\frac{3}{8}H^2,$$

$$\varkappa_+(H) = \frac{3}{8}\left[3\left(1 - \sqrt{\frac{5}{3}H^2}\right)^2 - H^2\right].$$

If $|H| < \sqrt{\frac{3}{5}}$, then the KL cycles are librating in the case of $\varkappa \in (\varkappa_0(H), \varkappa_+(H))$ and rotating in the case of $\varkappa \in [\varkappa_-(H), \varkappa_0(H))$. In the case of $\sqrt{\frac{3}{5}} < |H| < 1$, all the KL cycles are rotating with $\varkappa \in [\varkappa_-(H), \varkappa_0(H)]$.

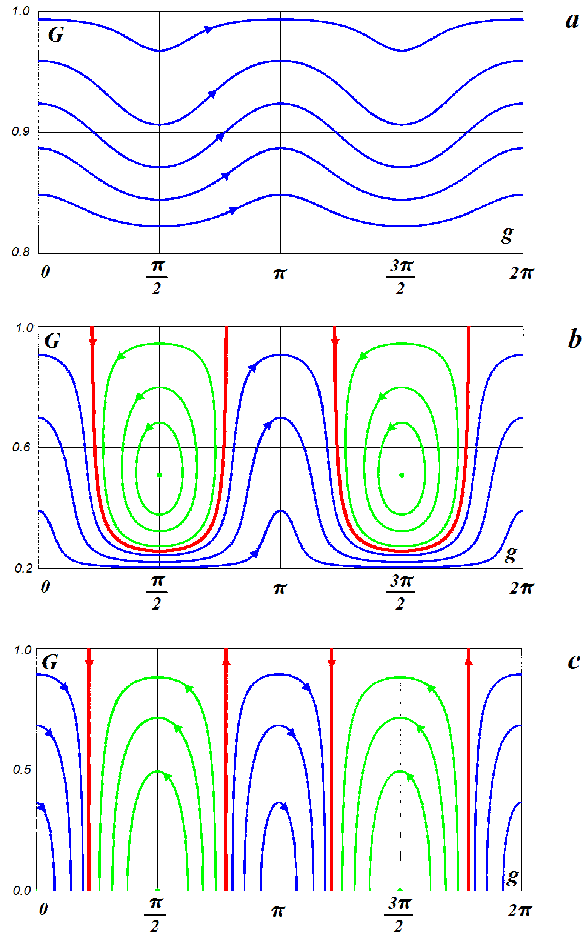

**Fig. 1** Phase portraits of a system with the Hamiltonian (2), with $H$ treated as parameter: a - $|H| = 0.8$, b - $|H| = 0.2$, c - $H = 0$. The trajectories depicted in green and blue correspond to librating and rotating KL cycles, respectively (or to their formal limits at $H = 0$). The separatrices (corresponding to aperiodic solutions) are shown in red.

The diagram in Figure 2 demonstrates the domains of the possible values of the integrals of motion in librating and rotating KL cycles in the halfplane $\varkappa, |H|$. This diagram is actually a well-known diagram from (Lidov 1962) redrawn in terms of other quantities. We prefer to depict it in the shown form, because it is in our plans to use a value of the Hamiltonian (2) as one of parameters defining the type of a KL cycle.

Explicit formulae describing the evolution of osculating elements in terms of elliptic functions can be found in (Gordeeva 1968; Kinoshita and Nakai 2007; Vashkovyak 1999) for both types of the KL cycles. Studying the EKL effect, we shall deal only with the relations characterising the rotating cycles.

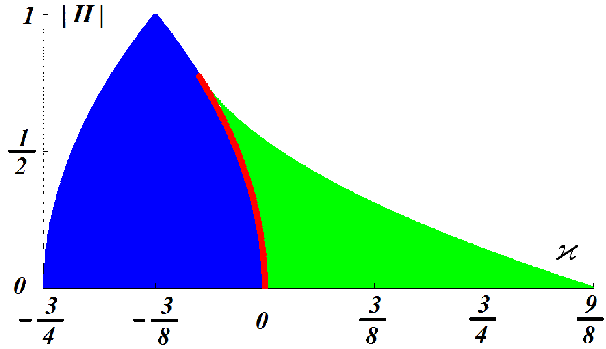

**Fig. 2** Connection of the modes of motion with the value of the quantities $\varkappa$, $H$. The domains of librating and rotating motions are shown in green and blue, respectively. The red line corresponds to aperiodic solutions.

In rotating KL cycles, the variable $g = \omega$ varies at a mean rate

$$n_g(\varkappa, H) = \frac{3\pi\sqrt{3(z_+ - z_-)}}{4\sqrt{2}\mathbf{K}(k_L)}, \quad (5)$$

where $\mathbf{K}(\cdot)$ is the complete elliptic integral of the first kind,

$$k_L^2 = \frac{z_0 - z_-}{z_+ - z_-}, \quad z_\pm = \frac{1}{2}\left(\chi \pm \sqrt{\chi^2 - \frac{20}{3}H^2}\right),$$

$$\chi = 1 - \frac{8}{9}\varkappa + \frac{4}{3}H^2, \quad z_0 = 1 + \frac{4}{3}\varkappa + \frac{1}{2}H^2.$$

Oscillations of the eccentricity and inclination in rotating KL cycles have the frequency $n_g(\varkappa, H)/2$. The mean rate of the ascending node precession is

$$n_h(\varkappa, H) = -\frac{3H}{4}\left[1 + \frac{2(z_0 - z_+)}{H^2 - z_+}\left(\frac{\mathbf{\Pi}(l_h^2, k_L)}{\mathbf{K}(k_L)} - 1\right)\right], \quad (6)$$

where $\mathbf{\Pi}(\cdot, \cdot)$ is the complete elliptic integral of the third kind,

$$l_h^2 = \frac{z_0 - z_-}{H^2 - z_-}.$$

The formulae (5) and (6) can be obtained from the related expressions for periods $T_g = 2\pi/n_g$ and $T_h = 2\pi/|n_h|$, presented in (Vashkovyak 1999). The reader, who wants to check it, must take into account that our choice of the unit of time differs from (Vashkovyak 1999) by a factor $3/16$.

2.2 Transition to "action-angle" variables

We start with the change of variables

$$(G, H, g, h) \mapsto (\widetilde{G}, \widetilde{H}, \widetilde{g}, \widetilde{h}), \tag{7}$$

where

$$\widetilde{G} = G - |H|, \ \widetilde{H} = H, \ \widetilde{g} = g, \ \widetilde{h} = h + \text{sign}(H) \cdot g.$$

It is easy to check that this change of variables is a canonical transformation with the generating function

$$S_1(\widetilde{G}, \widetilde{H}, g, h) = g(\widetilde{G} + |\widetilde{H}|) + h\widetilde{H}.$$

After the change of variables (7), the KL Hamiltonian

$$\widetilde{K}(\widetilde{G}, \widetilde{H}, \widetilde{g}, -) = K(\widetilde{G} + |\widetilde{H}|, \widetilde{H}, \widetilde{g}, -)$$

can still be considered as a Hamiltonian of the system with one degree of freedom depending on $\widetilde{H}$ as a parameter. Let $S_*(I_2, \widetilde{g}; \widetilde{H})$ be the generating function of a canonical transformation to the action-angle variables:

$$(G, g) \mapsto (I_2, \varphi_2),$$

where

$$I_2 = \frac{1}{2\pi} \int_{KL-cycle} \widetilde{G}(\varkappa, \widetilde{g}; \widetilde{H}) \, d\widetilde{g}. \tag{8}$$

The function $\widetilde{G}(\varkappa, \widetilde{g}; \widetilde{H})$ in (8) is the root of the equation [1]

$$\frac{4\varkappa}{3} + \frac{\widetilde{H}^2}{2} + \left[1 - (\widetilde{G} + |\widetilde{H}|)^2\right] \left\{1 - \frac{5\sin^2 \widetilde{g}}{2}\left[1 - \frac{\widetilde{H}^2}{(\widetilde{G} + |\widetilde{H}|)^2}\right]\right\} = 0, \tag{9}$$

satisfying the condition $0 \leq \widetilde{G} \leq 1 - |\widetilde{H}|$.

The relation (8) defines implicitly the function $\varkappa(\widetilde{H}, I_2)$. Then the expression for the generating function $S_*(I_2, \widetilde{g}; \widetilde{H})$ may be written as

$$S_*(I_2, \widetilde{g}; \widetilde{H}) = \int_0^{\widetilde{g}} \widetilde{G}(\varkappa(\widetilde{H}, I_2), \widetilde{g}'; \widetilde{H}) \, d\widetilde{g}'.$$

The transition to "action-angle" variables in the KL Hamiltonian

$$(\widetilde{G}, \widetilde{H}, \widetilde{g}, \widetilde{h}) \mapsto (I_1, I_2, \varphi_1, \varphi_2)$$

can now be carried out as a canonical transformation with the generating function

$$S_2(I_1, I_2, \widetilde{h}, \widetilde{g}) = I_1 \widetilde{h} + S_*(I_2, \widetilde{g}; I_1).$$

The relations between the variables $\widetilde{G}, \widetilde{H}, \widetilde{g}, \widetilde{h}$ and $I_1, I_2, \varphi_1, \varphi_2$ are

$$\widetilde{H} = I_1, \ \widetilde{G} = \frac{\partial S_*}{\partial \widetilde{g}}, \ \varphi_1 = \widetilde{h} + \frac{\partial S_*}{\partial I_1}, \ \varphi_2 = \frac{\partial S_*}{\partial I_2}. \tag{10}$$

---

[1] It should be noted that the equation (9) can be reduced to a biquadratic equation for $G = \widetilde{G} + |\widetilde{H}|$.

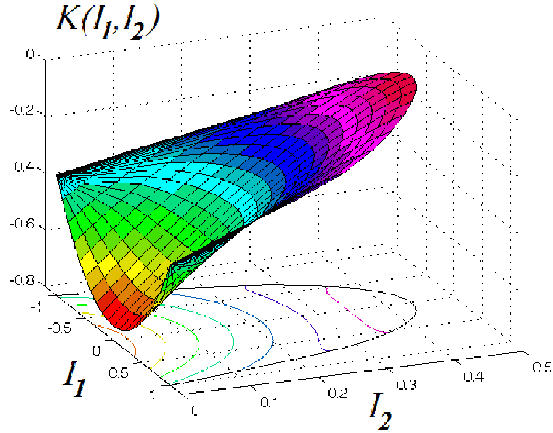

**Fig. 3** The KL Hamiltonian as a function of the variables $I_1, I_2$

After the transition to the variables $I_1, I_2, \varphi_1, \varphi_2$, the KL Hamiltonian assumes the form of

$$K(I_1, I_2) = \varkappa(I_1, I_2).$$

In the plane $I_1, I_2$, the domain of definition of the function $K(I_1, I_2)$ consists of the points satisfying the condition

$$|I_1| \leq B(I_2),$$

where the function $I_1 = B(I_2)$ is defined implicitly by the relation $\varkappa_0(I_1) = \varkappa(I_1, I_2)$.

A 3D-graph of the function $K(I_1, I_2)$ is shown in Figure 3. It was constructed by numerical inversion of the relation (8) with $\widetilde{H}$ replaced by $I_1$.

Within the classical Kozai-Lidov model ($\varepsilon = 0$), the angle variables $\varphi_1, \varphi_2$ vary, in general case, as linear functions of time:

$$\varphi_1 = n_L(\varkappa, I_1)t + \varphi_{10}, \quad \varphi_2 = n_g(\varkappa, I_1)t + \varphi_{20}.$$

Here $\varphi_{10}, \varphi_{20}$ are arbitrary constants,

$$n_L(\varkappa, I_1) = n_h(\varkappa, I_1) + \operatorname{sign} I_1 \cdot n_g(\varkappa, I_1) = \quad (11)$$

$$= -\frac{I_1}{2}\left[\frac{5c_*}{1 - z_- - \frac{2}{3}c_*}\frac{\mathbf{\Pi}(l_L^2, k_L)}{\mathbf{K}(k_L)} + \frac{3}{2}\right],$$

$$l_L^2 = \frac{k_L^2}{l_h^2} = \frac{z_0 - z_-}{1 - z_- - \frac{2}{3}c_*}, \quad c_* = \frac{4}{3}\varkappa + \frac{H^2}{2}.$$

To obtain the expression (11) for $n_L(\varkappa, I_1)$, the relation

$$\mathbf{\Pi}(l_L^2, k_L) = -\mathbf{\Pi}(l_h^2, k_L) + \mathbf{K}(k_L) + \frac{\pi}{2}\sqrt{\frac{l_h^2}{(1 - l_h^2)(l_h^2 - k_L^2)}}$$

was applied. This relation is based on formula (117.02) from (Byrd and Friedman 1954).

The index "L" for the rate of variation of the variable $\varphi_1$ is used to emphasize the geometric meaning of this variable: it characterizes the secular component in the variation of the longitude $\Omega_L$ of the Laplace vector directed to the pericentre of the test particle's osculating orbit. More precisely, one can show that

$$\Omega_L = \varphi_1 + \Omega_L^*(\varphi_2, I_1, I_2), \tag{12}$$

where the second term in (12) describes the periodic component which depends on the parameters of the KL cycle:

$$\Omega_L^*(\varphi_2 + \pi k, I_1, I_2) = \Omega_L^*(\varphi_2, I_1, I_2), \ k \in Z^1,$$

$$\Omega_L^*(-\varphi_2, I_1, I_2) = -\Omega_L^*(\varphi_2, I_1, I_2), \quad \Omega_L^*(\varphi_2, I_1, I_2) = O(I_1).$$

The formula (12) is derived in Appendix A.

The case of $I_1 = 0$ corresponds to the motion of the test particle in the orbit with the inclination $i = 90°$. Since $n_L(\varkappa, 0) = 0$, we can interpret this case as a resonance. This will enable us to apply the standard methods for investigation of resonances in Hamiltonian systems (see, e.g., Arnold et al. (2006)), in order to study the dynamics of the perturbed system (1) for $|I_1| \lesssim \varepsilon^{1/2}$. This rough estimate is related to a non-degenerate situation; otherwise a special consideration is required to determine the size of the resonance zone.

2.3 Expansion of the KL Hamiltonian over the powers of $I_1$

Since the value of the "action" $I_1$ will hereafter be assumed small, it is natural to expand the KL Hamiltonian $K(I_1, I_2)$ to a power series in this variable:

$$K(I_1, I_2) = K_0(I_2) + K_2(I_2) \cdot I_1^2 + K_4(I_2) \cdot I_1^4 + \ldots \tag{13}$$

To find the first coefficients of the series (13), we need several auxiliary relations. To start with, we write down the expression for the integral (8) in the form of a series:

$$I_2(\varkappa, I_1) = g_0(\varkappa) + g_2(\varkappa) \cdot I_1^2 + \ldots \tag{14}$$

Here, as before, $\varkappa$ denotes the value of the KL Hamiltonian for a corresponding KL cycle. The first term in (14) is equal to the integral (8) over the degenerate cycles in which the inclination of the particle has a fixed value $90°$:

$$g_0(\varkappa) = \frac{1}{\pi} \sqrt{\frac{10}{1-\alpha}} \left[ \mathbf{\Pi}\left(\frac{\alpha}{1-\alpha}, k\right) - (1-\alpha) \mathbf{K}(k) \right], \tag{15}$$

with $\alpha$ given by

$$\alpha = \frac{2}{5}\left(1 + \frac{4}{3}\varkappa\right), \ k^2 = \frac{3\alpha}{2(1-\alpha)} = \frac{3(3+4\varkappa)}{9-8\varkappa}.$$

Note that degenerate rotating KL cycles exist for $\varkappa \in [-3/4, 0]$. It is easy to see that

$$g_0\left(-\frac{3}{4}\right) = 0, \ g_0(0) = \frac{2}{\pi} \arcsin\sqrt{\frac{2}{5}} \approx 0.435906.$$

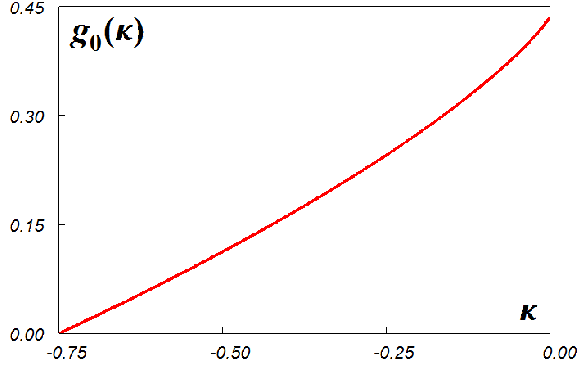

**Fig. 4** The graph of the function $g_0(\varkappa)$

The graph of the function (15) is shown in Figure 4.

Inverting the function $g_0(\varkappa)$, we obtain the first coefficient $K_0(I_2)$ of the series (13):

$$K_0(g_0(\varkappa)) = \varkappa, \ g_0(K_0(I_2)) = I_2.$$

To find the next coefficients of the series (13), we expand $n_L(\varkappa, I_1)$ into a series in $I_1$:

$$n_L(\varkappa, I_1) = f_1(\varkappa) \cdot I_1 + f_3(\varkappa) \cdot I_1^3 + \ldots \qquad (16)$$

where

$$f_1(\varkappa) = \frac{3}{2}\left(\frac{\mathbf{E}(k)}{\mathbf{K}(k)} - \frac{1}{2}\right),$$

$$f_3(\varkappa) = \frac{45}{8(9-8\varkappa)^3}\left\{-\frac{81 + 336\varkappa + 64\varkappa^2}{k^2} + \right.$$

$$+\left[\frac{81 + 432\varkappa + 192\varkappa^2}{k^2} - \frac{81 + 240\varkappa - 64\varkappa^2}{k'^2} + \frac{81 + 336\varkappa + 64\varkappa^2}{k^2 k'^2}\right]\frac{\mathbf{E}(k)}{\mathbf{K}(k)} -$$

$$\left. -\frac{3(27 + 144\varkappa + 64\varkappa^2)}{k^2 k'^2}\left(\frac{\mathbf{E}(k)}{\mathbf{K}(k)}\right)^2\right\}.$$

Now let us take into account the evident relation

$$\frac{\partial K}{\partial I_1} = n_L(K(I_1, I_2), I_1). \qquad (17)$$

Substitution of the series (13) and (16) into the formula (17) entails the following result:

$$K_2(I_2) = \frac{1}{2}f_1(K_0(I_2)),$$

$$K_4(I_2) = \frac{1}{4}\left(f_3(K_0(I_2)) + \frac{f_1(K_0(I_2))}{2} \cdot \left.\frac{\partial f_1}{\partial \varkappa}\right|_{\varkappa = K_0(I_2)}\right),$$

where
$$\frac{\partial f_1}{\partial \varkappa} = -\frac{9}{4\varkappa(3+4\varkappa)}\left[k'^2 - 2k'^2\left(\frac{\mathbf{E}(k)}{\mathbf{K}(k)}\right) + \left(\frac{\mathbf{E}(k)}{\mathbf{K}(k)}\right)^2\right].$$

The graphs of the functions $K_0(I_2)$, $K_2(I_2)$ and $K_4(I_2)$ are shown in Figure 5. The domain of the definition of these functions is the interval $[0, I_2^{max}]$, where $I_2^{max} = g_0(0)$.

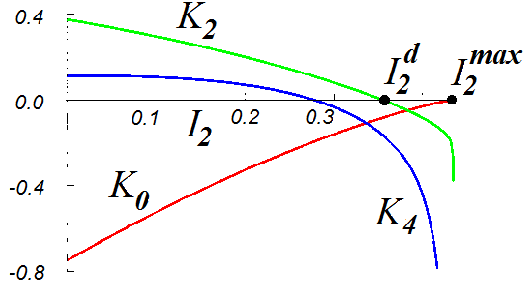

**Fig. 5** Graphs of the functions $K_0(I_2), K_2(I_2), K_4(I_2)$

In Figure 6 we present the graph of the function $\Delta(I_1, I_2)$ characterizing the difference between $K(I_1, I_2)$ and its approximation by the sum of the first three terms of (13). This function weakly depends on $I_2$. For $|I_1| < 0.5$ the value of $\Delta(I_1, I_2)$ is smaller $0.01$.

As will be established in Section 4, a specific situation arises in the analysis of the EKL effect when $I_2 \approx I_2^d$. In that case, $I_2^d = 0.35627\ldots$ is the root of the equation $K_2(I_2) = 0$. It is easy to find that

$$\left.\frac{\partial K_2}{\partial I_2}\right|_{I_2=I_2^d} = \frac{n_g(\varkappa,0)}{2} \cdot \left.\frac{\partial f_1}{\partial \varkappa}\right|_{\varkappa=K_0(I_2^d)} \approx -1.61832,$$

$$K_4(I_2^d) \approx -0.16324.$$

The minimal and maximal values of the eccentricity and inclination for rotating KL cycles with $0 < I_1 \ll 1$ can be calculated using the formulae

$$e_{min}^2 \approx -\frac{4}{3}K_0(I_2) + O(I_1^2), \tag{18}$$

$$e_{max}^2 \approx 1 - \frac{15 I_1^2}{9 - 8K_0(I_2)} + O(I_1^4).$$

The minimal and maximal values of the inclination for these cycles are

$$i_{min} = \arccos\frac{I_1}{\sqrt{1-e_{max}^2}}, \quad i_{max} = \arccos\frac{I_1}{\sqrt{1-e_{min}^2}}$$

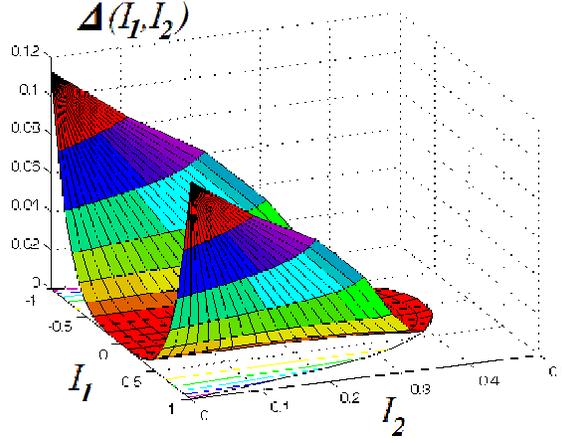

**Fig. 6** The error of the approximation of $K(I_1, I_2)$ by the sum of the first three terms of (13)

in the case $I_1 > 0$ and

$$i_{min} = \pi - \arccos \frac{I_1}{\sqrt{1-e_{min}^2}}, \; i_{max} = \pi - \arccos \frac{I_1}{\sqrt{1-e_{max}^2}}$$

in the case $I_1 < 0$.

## 3 Analysis of the secular effects due to the octupole component of the disturbing function. The first-order perturbation theory

3.1 Averaging over KL cycles

After the transition to the "action-angle" variables, the Hamiltonian (1) takes the form

$$\mathcal{K}(I_1, I_2, \varphi_1, \varphi_2) = K(I_1, I_2) + \varepsilon K^*(I_1, I_2, \varphi_1, \varphi_2).$$

As a consequence of the symmetry properties (3) we have

$$\mathcal{K}(-I_1, I_2, 2\pi - \varphi_1, \varphi_2) = \mathcal{K}(-I_1, I_2, \varphi_1, 2\pi - \varphi_2) = \mathcal{K}(I_1, I_2, \varphi_1, \varphi_2). \quad (19)$$

Like in the other cases, we expand the perturbing term $K^*(I_1, I_2, \varphi_1, \varphi_2)$ to a power series in $I_1$:

$$K^*(I_1, I_2, \varphi_1, \varphi_2) = K_0^*(I_2, \varphi_1, \varphi_2) + I_1 \cdot K_1^*(I_2, \varphi_1, \varphi_2) + \ldots$$

The relations (19) imply

$$K_{2j-1}^*(I_2, \varphi_1, \varphi_2) = -K_{2j-1}^*(I_2, \varphi_1, 2\pi - \varphi_2), \quad (20)$$

$$K_{2(j-1)}^*(I_2, \varphi_1, \varphi_2) = K_{2(j-1)}^*(I_2, \varphi_1, 2\pi - \varphi_2), \quad j = 1, 2 \ldots.$$

When "action-angle" variables are introduced, the averaging over a KL cycle reduces to averaging over the variable $\varphi_2$. Let us introduce the notation

$$\langle \cdot \rangle_{\varphi_2} = \frac{1}{2\pi} \int_0^{2\pi} (\cdot)\, d\varphi_2.$$

Taking into account the relations (20), we obtain:

$$\langle K^*_{2j-1} \rangle_{\varphi_2} = 0,\ j = 1, 2, \ldots$$

Thus the result of averaging of $K^*(I_1, I_2, \varphi_1, \varphi_2)$ can be written as

$$\langle K^* \rangle_{\varphi_2} = \langle K_0^* \rangle_{\varphi_2} + O(I_1^2). \tag{21}$$

The calculation of the principal term on the right-hand side of the relation (21) implies averaging along the motions with the inclination $i = 90°$. Carried out by Katz et al. (2011), this step renders:

$$\langle K_0^* \rangle_{\varphi_2} = -\cos\varphi_1 \cdot Q(I_2),$$

where

$$Q(I_2) = \left. \frac{5\pi(3+11\varkappa)\sqrt{9-8\varkappa}}{32\sqrt{15}\mathbf{K}(k)} \right|_{\varkappa = K_0(I_2)}.$$

The graph of the function $Q(I_2)$ is presented in Figure 7. This function equals zero for $I_2 = I_2^* \approx 0.2321$ and $I_2 = I_2^{max}$. When $I_2$ tends to $I_2^{max}$

$$Q(I_2) \approx \frac{3\pi\sqrt{15}}{16 \ln \frac{36}{5|K_0(I_2)|}}.$$

Also note that

$$Q(0) = -\frac{105}{64},\ \frac{dQ}{dI_2} = \frac{\pi}{32\mathbf{K}(k)} \sqrt{\frac{15}{9-8\varkappa}} [29 - 44\varkappa +$$

$$\left. \frac{30(3+11\varkappa)}{k^2 k'^2(9-8\varkappa)} \left(1 - \frac{\mathbf{E}(k)}{\mathbf{K}(k)}\right)\right]_{\varkappa=K_0(I_2)} \cdot n_g(K_0(I_2), 0).$$

3.2 Evolution of the motion in the nondegenerate case

In the first approximation of the perturbation theory, after averaging over a KL cycle the evolution of the variables $I_1, \varphi_1$ is described by the 1DOF Hamiltonian system

$$\mathcal{L}(I_1, \varphi_1) = K_2(I_2) \cdot I_1^2 + K_4(I_2) \cdot I_1^4 - \varepsilon \cos\varphi_1 \cdot Q(I_2) + O(I_1^6, \varepsilon I_1^2), \tag{22}$$

where $I_2$ plays the role of a fixed parameter.

Investigation of the secular effects will begin with the case $K_2(I_2) \neq 0$, $Q(I_2) \neq 0$. Following the standard approach to the analysis of resonances in Hamiltonian systems, we re-scale the variable $I_1$ and the independent variable $t$:

$$I_1 \mapsto J_1 = \varepsilon^{-1/2} I_1,\ t \mapsto t^* = \varepsilon^{1/2} t.$$

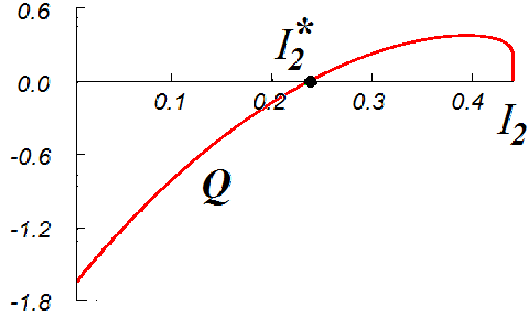

**Fig. 7** The graph of the function $Q(I_2)$

Omitting the terms of a higher order of smallness, we find that the evolution of the variables $J_1, \varphi_1$ is described by a Hamiltonian system of the pendulum type:

$$\mathcal{L}^*(J_1, \varphi_1) = K_2(I_2) \cdot J_1^2 - \cos\varphi_1 \cdot Q(I_2). \qquad (23)$$

A system with the Hamiltonian (23) has two families of equilibrium solutions:

$$J_1 = 0, \ \varphi_1 = 0 \ (\text{mod } 2\pi) \qquad (24)$$

and

$$J_1 = 0, \ \varphi_1 = \pi \ (\text{mod } 2\pi). \qquad (25)$$

If $I_2 \in (I_2^*, I_2^d)$, then the equilibria (24) are stable, whereas the equilibria (25) are unstable. In the case of $I_2 \in [0, I_2^*) \cup (I_2^d, I_2^{max})$, the situation is opposite: the equilibria (25) are stable, whereas the equilibria (24) are not.

The change in the sign of $J_1$ in oscillating periodic solutions of the system (23) corresponds to a change in the direction of the orbital motion of the test particle. So the consideration of (23) allows us to make some qualitative and quantitative conclusions regarding the properties of the EKL effect.

To begin with, we estimate the characteristic time between flips $T_{EKL}$. The simplest way to do it is to compute the value of a semi-period of small oscillations around the stable equilibria. In the original variables we obtain

$$T_{EKL} = \frac{\pi}{\varepsilon^{1/2}\sqrt{2|K_2(I_2)Q(I_2)|}}.$$

A similar asymptotics was obtained by Antognini (2015), however the fact, that under certain conditions $T_{EKL}$ dependence on $\varepsilon$ differs, was absent in his argumentation (see Sections 4 and 5).

The upper limit of the possible absolute values of the variable $I_1$ in motion with flips is given by the formula

$$I_1^{sup} = \varepsilon^{1/2}\sqrt{2\left|\frac{Q(I_2)}{K_2(I_2)}\right|}. \qquad (26)$$

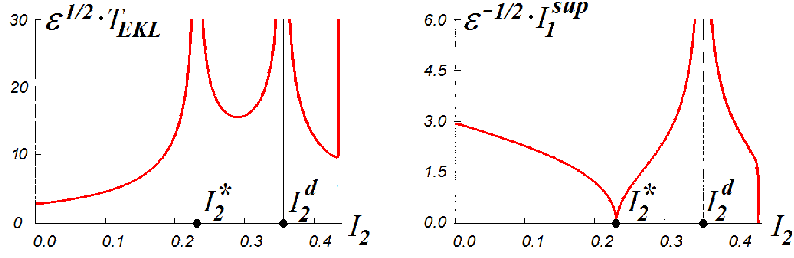

**Fig. 8** The dependence of the characteristic time of the EKL effect and of the maximum absolute values of the variable $I_1$ in motions with flips on the value of the variable $I_2$

In Figure 8, we presented graphs illustrating the dependence of $T_{EKL}$ and $I_1^{sup}$ on the value of the variable $I_2$ in the corresponding motion. Singularities at $I_2 = I_2^*$, $I_2 = I_2^d$ and $I_2 = I_2^{max}$ indicate the another order of asymptotics in these situations. The cases of $I_2 \approx I_2^*$ and $I_2 \approx I_2^d$ will be considered in Sections 4 and 5, respectively; the case of $I_2 \approx I_2^{max}$ takes place in a relatively small region of the phase space and will not be addressed in detail.

The developed theory is valid in case

$$I_1^{sup} \ll B(I_2), \qquad (27)$$

where the function $B(I_2)$ was introduced in Section 2.2 (in other words, the resonance region must be within the interval of possible values of $I_1$ for given value of the "action" variable $I_2$). For moderate values of $\varepsilon$ (typically $0.1 \div 0.01$), when the condition (27) is not satisfied, the system demonstrates chaotic behavior. In (Li et al. 2014a) the contribution of overlapping secondary resonances in the formation of chaotic dynamics was considered.

## 4 Secular evolution in the case of $I_2 \approx I_2^d$

4.1 An approximate Hamiltonian comprising the leading terms

We begin with a rescaling transformation

$$I_1 \mapsto \hat{J}_1 = \varepsilon^{-1/4} I_1, \ I_2 \mapsto \hat{J}_2 = \varepsilon^{-1/2}(I_2 - I_2^d), \ t \mapsto \hat{t} = \varepsilon^{3/4} t. \qquad (28)$$

Evolution of the variables $\hat{J}_1, \varphi_1$ is described by the approximate equations

$$\frac{d\hat{J}_1}{d\hat{t}} = -\frac{\partial \hat{\mathcal{L}}}{\partial \varphi_1}, \ \frac{d\varphi_1}{d\hat{t}} = \frac{\partial \hat{\mathcal{L}}}{\partial \hat{J}_1}, \qquad (29)$$

where

$$\hat{\mathcal{L}}(\hat{J}_1, \varphi_1) = \sigma \hat{J}_1^2 + \hat{K}_4 \hat{J}_1^2 - \hat{Q} \cos \varphi_1,$$

$$\hat{K}_4 = K_4(I_2^d), \ \hat{Q} = Q(I_2^d), \ \sigma = \hat{J}_2 \left.\frac{dK_2}{dI_2}\right|_{I_2=I_2^d}.$$

Qualitative behaviour of the solutions to the system (29) depends on the value of the parameter $\sigma$. In the case of $\sigma < 0$, the phase portrait of the system (29) is topologically equivalent to the phase portrait of a mathematical pendulum (Figure 9a). Let $S$ denote a family of oscillatory solutions to (29). With this family, we associate the quantity $\hat{J}_1^{sup}$, which limits the absolute value of the variable $\hat{J}_1$ in oscillating solutions:

$$\hat{J}_1^{sup} = \sup_{(\hat{J}_1(\hat{t}), \varphi_1(\hat{t})) \in S, \hat{t} \in R^1} \left| \hat{J}_1(\hat{t}) \right|.$$

In the considered case, it is easy to find that

$$\hat{J}_1^{sup} = \sqrt{\frac{\sqrt{\sigma^2 - 8\hat{Q}\hat{K}_4} + \sigma}{2\left|\hat{K}_4\right|}}. \tag{30}$$

When $\sigma = 0$, bifurcation takes place. It leads to the emergence of an equilibrium solutions residing outside the axis $\hat{J}_1 = 0$ for positive $\sigma$. Stable and unstable solutions of this type are given by the relations

$$\hat{J}_1 \equiv \pm \hat{J}_1^d, \; \varphi_1 \equiv 0 \,(\mathrm{mod}\, 2\pi)$$

and

$$\hat{J}_1 \equiv \pm \hat{J}_1^d, \; \varphi_1 \equiv \pi \,(\mathrm{mod}\, 2\pi)$$

respectively, where

$$\hat{J}_1^d = \sqrt{\frac{\sigma}{2\left|\hat{K}_4\right|}}.$$

If $0 < \sigma < \sigma_d = \sqrt{8\hat{Q}\left|\hat{K}_4\right|} \approx 0.67837$, then there exist oscillating solutions with the phase trajectories enclosing two stable equilibria (Figure 9b). For such solutions,

$$\hat{J}_1^{sup} = \sqrt{\frac{\sigma + \sqrt{8\hat{Q}\left|\hat{K}_4\right|}}{2\left|\hat{K}_4\right|}}. \tag{31}$$

These solutions disappear as a result of another bifurcation at $\sigma = \sigma_d$. Figure 9c presents an example of the phase portrait of the system (29) for $\sigma > \sigma_d$. In this case,

$$\hat{J}_1^{sup} = \sqrt{\frac{\sigma - \sqrt{\sigma^2 + 8\hat{Q}\hat{K}_4}}{2\left|\hat{K}_4\right|}}. \tag{32}$$

The graph of the function $\hat{J}_1^{sup}(\sigma)$ defined by formulae (30),(31) and (32) is shown in Figure 10.

The discussed bifurcations are well known in the general theory of Hamiltonian systems (e.g., Howard and Humpherys (1995); Morozov (2002)).

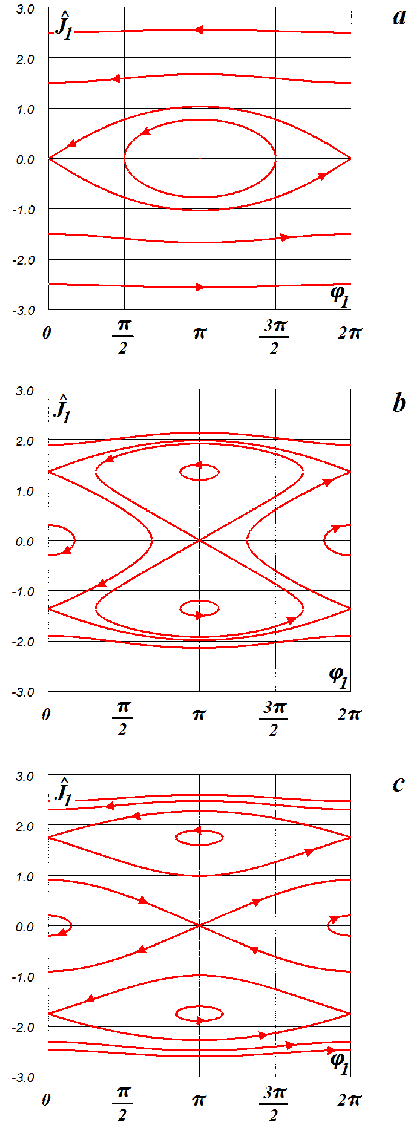

**Fig. 9** Phase portraits of the system (29) for different values of the parameter $\sigma$: a - $\sigma = -0.5$, b - $\sigma = 0.6$, c - $\sigma = 1.0$

4.2 Interpretation in terms of Keplerian elements

The results of Section 4.1 allow us to establish the upper limit of the possible values of $|I_1|$ in motions with flips, when the value of $I_2$ is close to that of $I_2^d$:

$$I_1^{\text{sup}} \approx \varepsilon^{1/4} \sqrt{\frac{8\hat{Q}}{\left|\hat{K}_4\right|}}. \tag{33}$$

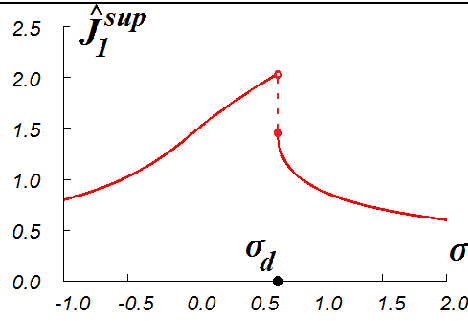

**Fig. 10** The graph of the function $\hat{J}_1^{sup}$

More precisely, the amplitude of the variation of the variable $I_1$ reaches its maximal value (33) at

$$I_2 \approx I_2^d + \frac{\varepsilon^{1/2}\sigma_d}{dK_2/dI_2(I_2^d)}.$$

A typical time between flips in such motions is $T_{EKL} \sim \varepsilon^{-3/4}$. It is significantly longer than in the case $|I_2 - I_2^d| \sim 1$ studied in Section 3.

Lithwick and Naoz (2011) numerically constructed curves on the plane of initial values $(e_0, \cos i_0)$, separating for different values of $\varepsilon$ the motions with flips and the motions without flips. The graph of the function $\hat{J}_1^{\text{sup}}(\sigma)$, complemented by a segment at the point of discontinuity, can be interpreted as a similar graph in the plane of rescaled variables. Its "image" in the plane $(e_0, \cos i_0)$ provides an approximate analytical expression of the separating curves:

$$e_0(\sigma) = e_d - \frac{\varepsilon^{1/2}}{\sqrt{3\left|\hat{K}_0\right|}} \left[\frac{3}{8}\left(\hat{J}_1^{\text{sup}}(\sigma)\right)^2 + \sigma\frac{dK_0/dI_2(I_2^d)}{dK_2/dI_2(I_2^d)}\right], \qquad (34)$$

$$\cos i_0(\sigma) = \frac{\varepsilon^{1/4}\hat{J}_1^{\text{sup}}(\sigma)}{\sqrt{1-e_d^2}}.$$

where

$$e_d = 2\sqrt{\frac{\left|\hat{K}_0\right|}{3}}.$$

Good agreement between the asymptotic formulae (34) and the results of numerical simulations is achieved only for very small values of the parameter $\varepsilon$ (Fig. 11). However, the qualitative properties of the motion described in this section (in particular, the specific bifurcations defining the possible scenarios of the long-term evolution) hold even when the formulae (34) are inapplicable ($\varepsilon \sim 0.1 \div 0.01$)

To obtain similar expression for the separating curve in the case of $|I_2 - I_2^d| \sim 1$, one can substitute $I_1^{sup}(I_2)$ for $I_1$ (where $I_1^{sup}(I_2)$ is given by the formula (26)) in the estimates (18). This yields an approximate parametric representation (with $I_2$ as a parameter) of the curve separating on the plane $(e_0, \cos i_0)$ the initial values

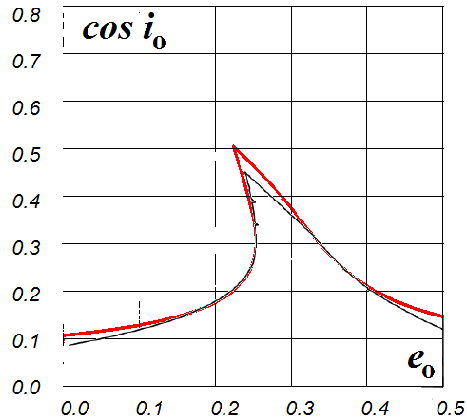

**Fig. 11** The red curve is defined by our equation (34). In black is shown the separating curve constructed numerically by Lithwick and Naoz (2011) for $\varepsilon = 0.003$

for motion with and without flips. Of special interest is the situation with $I_2 \ll 1$ ($\varkappa \approx -4/3$), since it corresponds to the "low-inclination" EKL effect discussed by Li et al. (2014b)

## 5 The degenerate case of $Q(I_2) \approx 0$

The degenerate case of $Q(I_2) \approx 0$ takes place at $I_2 \approx I_2^*$ or $I_2 \approx I_2^{\max}$. This is a situation where a second-order perturbation theory is required to reveal the character of the long-term evolution of the motion. For the model to be correct, it will be necessary to take into account the next, hexadecapole, term in the expansion of the disturbing function. Without carrying out these (rather cumbersome) calculations, we would only note that, for the degeneracy under consideration, a typical time between flips obeys $T_{EKL} \sim \varepsilon^{-1} \gg \varepsilon^{-1/2}$, while the upper limit of the absolute values of the variable $I_1$ in the motions with flips is proportional to $\varepsilon$.

## 6 Conclusions

We have demonstrated a possibility to interpret the EKL effect as a resonance phenomenon. This has allowed us to obtain the previously known properties of this effect in a more straightforward way and to establish new properties thereof.

To reduce the Hamiltonian to a form conventionally used in the modern theory of resonances in Hamiltonian systems, we introduced the action-angle variables in the classical Kozai-Lidov model of weakly perturbed Keplerian motion of the test particle. Analysis of the general case, and of various special cases with the Hamiltonian being degenerate in certain sense, has enabled us to provide a more accurate estimate for the time interval $T_{EKL}$ between flips (or, stated differently, between changes in the direction of the test particle's orbital motion). The interval $T_{EKL}$ turns out to be of the order of $\varepsilon$ taken to some power. Dependent on the initial conditions, we may obtain either $T_{EKL} \sim \varepsilon^{-1/2}$, or $T_{EKL} \sim \varepsilon^{-3/4}$, or $T_{EKL} \sim \varepsilon^{-1}$. Here $\varepsilon$ is a small parameter characterising the importance of the octupole term in the expansion of the disturbing function, while the units of time are chosen so that the period of a KL cycle $T_{KL} \sim 1$.

We also have discovered the bifurcations governing the possible scenarios of the secular evolution, and have derived an asymptotic formula for the boundary curve separating the initial conditions giving rise to motions with and without flips.

**Acknowledgments**

The work was supported by the Presidium of the Russian Academy of Sciences (Program 7 "Experimental and theoretical studies of the objects in the Solar system and exoplanetary systems"). We are grateful to S.Breiter, S.S.Efimov, M.Efroimsky, S.Naoz, A.I.Neishtadt, D.A.Pritykin, and M.A.Vashkovyak for reading the manuscript and useful discussions. We appreciate A.Rosengren's friendly consultation on scientific writing in English. We also thank anonymous referees for all their corrections and suggestions.

**Appendix A. Evolution of the Laplace vector in rotating KL-cycles**

The evolution of the Laplace vector

$$\mathbf{e}_L = e \left(\cos\vartheta_L \cos\Omega_L, \cos\vartheta_L \sin\Omega_L, sin\vartheta_L\right)^T$$

is described by the relations characterizing the change in eccentricity $e$ of the orbit, the latitude $\vartheta_L$ and the longitude $\Omega_L$ of this vector.

Formulae determining the change in eccentricity in KL-cycles are given in (Gordeeva 1968; Vashkovyak 1999). In our notation the formula for the rotating cycles takes the form:

$$e(t) = \sqrt{1 - z_- - (z_+ - z_-)\operatorname{sn}^2(u(t), k_L)}, \qquad (35)$$

$$u(t) = \frac{3}{2}\sqrt{\frac{3(z_+ - z_-)}{2}} t + u_0.$$

To find the latitude $\vartheta_L$ of the Laplace vector, we use the relation

$$\sin\vartheta_L = \sin i \sin\omega.$$

Writing the Kozai-Lidov Hamiltonian in form

$$K = -\frac{3}{8}\left[H^2 + 2e^2\left(1 - \frac{5}{2}\sin^2 i \sin^2\omega\right)\right],$$

it is easy to establish that

$$\sin^2\vartheta_L = \frac{2}{5}\left(1 + \frac{c_*}{e^2}\right). \qquad (36)$$

The change in the longitude of the Laplace vector is described by the differential equation

$$\dot{\Omega}_L = 3H\left(1 - \frac{3}{4\cos^2\vartheta_L}\right). \tag{37}$$

Using (35) and (36), we can transform (37) to the form

$$\dot{\Omega}_L = -\frac{H}{2}\left[\frac{3}{2} + \frac{5c_*}{\left(1 - z_- - \frac{2}{3}c_*\right)} \cdot \frac{1}{\left(1 - l_L^2\,\mathrm{sn}^2(u(t), k_L)\right)}\right]. \tag{38}$$

Integrating (38), we get

$$\Omega_L(t) = \Omega_{L0} - \frac{H}{2}\left[\frac{3}{2}t + \frac{5\sqrt{2}c_*}{3\sqrt{3(z_+ - z_-)}\left(1 - z_- - \frac{2}{3}c_*\right)} \int_{u_0}^{u(t)} \frac{du'}{\left(1 - l_L^2\,\mathrm{sn}^2(u', k_L)\right)}\right]. \tag{39}$$

For the next step we need the relation

$$\int_0^u \frac{du'}{\left(1 - l_L^2\,\mathrm{sn}^2(u', k_L)\right)} = \frac{\mathbf{\Pi}(l_L^2, k_L)}{\mathbf{K}(k_L)}u + \Phi_L(u), \tag{40}$$

where

$$\Phi_L(u) = \Pi(\mathrm{am}(u), l_L^2, k_L) - \frac{\mathbf{\Pi}(l_L^2, k_L)}{\mathbf{K}(k_L)}u.$$

It is worth noting that $\Phi_L(u)$ is an odd periodic function of $u$ with the period $2K(k_L)$.

From (39) and (40) it follows, that $\Omega_L(t)$ can be represented as the sum of a linear function of time $\overline{\Omega}(t)$, characterizing the secular evolution of the longitude of the Laplace vector and $(T_g/2)$-periodic function of time $\widetilde{\Omega}(t)$:

$$\Omega_L(t) = \overline{\Omega}_L(t) + \widetilde{\Omega}_L(t), \tag{41}$$

$$\widetilde{\Omega}_L(t) = -\frac{5\sqrt{2}c_*}{3\sqrt{3(z_+ - z_-)}\left(1 - z_- - \frac{2}{3}c_*\right)}\Phi_L(u(t)),$$

$$\overline{\Omega}_L(t) = n_L t + \Omega_{L0} - \widetilde{\Omega}_L(0).$$

Having the relation (41), it is not too difficult to reconstruct the expression for $\Omega_L$ as a function of "action-angle" variables. In that case when $\omega = \varphi_2 = 0(\mathrm{mod}\,2\pi)$, the eccentricity $e$ in rotating KL-cycles has a minimal value. From the formula (35) it follows that at these moments of time

$$u = K(k_L)(\mathrm{mod}\,2K(k_L)).$$

Thus, the values of $u$ and $\varphi_2$ are related by the equality

$$u = 2K(k_L)\left(\frac{\varphi_2}{\pi} + \frac{1}{2}\right)(\mathrm{mod}\,2K(k_L)). \tag{42}$$

Let $t_j$ $(j \in Z^1)$ denote the moment of time when

$$\omega = \varphi_2 = 0(\mathrm{mod}\,2\pi). \tag{43}$$

Substituting $u = K(k_L)(\mathrm{mod}\, 2K(k_L))$ into the formula for $\widetilde{\Omega}(t)$, we obtain:

$$\widetilde{\Omega}(t_j) = 0. \tag{44}$$

On the other hand, as a consequence of the third relation in (10), in the case (43) we have

$$\Omega_L = \Omega = \varphi_1 (\mathrm{mod}\, 2\pi).$$

If we take into account (41) and (44), then we get at time moments $t_j$

$$\widetilde{\Omega} = \varphi_1 (\mathrm{mod}\, 2\pi). \tag{45}$$

Since the rates of change of $\overline{\Omega}$ and $\varphi_1$ are equal, the relation (45) is valid at any other moment of time.

From the relations (41),(42) and (44) follows that

$$\Omega_L = \varphi_1 + \Omega_L^*(\varphi_2, I_1, I_2),$$

where

$$\Omega_L^*(\varphi_2, I_1, I_2) = \frac{5 I_1 c_*}{3\left(1 - z_- - \frac{2}{3} c_*\right)} \sqrt{\frac{2}{3(z_+ - z_-)}} \Phi_L\left(\frac{2\mathbf{K}(k_L)\varphi_2}{\pi}\right),$$

**Appendix B. Some KAM-theory-based conclusions on the properties of motions with small absolute values of the variable $I_1$**

We recall that after the transition to the "action-angle" variables the Hamiltonian of the problem acquires the form

$$\mathcal{K}(I_1, I_2, \varphi_1, \varphi_2) = K(I_1, I_2) + \varepsilon K^*(I_1, I_2, \varphi_1, \varphi_2). \tag{46}$$

For Hamiltonians of the form (46), it is established in the KAM theory that under the condition of isoenergetic nondegeneracy,

$$D(I_1, I_2) = \begin{vmatrix} \frac{\partial^2 K}{\partial I_1^2} & \frac{\partial^2 K}{\partial I_1 \partial I_2} & \frac{\partial K}{\partial I_1} \\ \frac{\partial^2 K}{\partial I_1 \partial I_2} & \frac{\partial^2 K}{\partial I_2^2} & \frac{\partial K}{\partial I_2} \\ \frac{\partial K}{\partial I_1} & \frac{\partial K}{\partial I_2} & 0 \end{vmatrix} \neq 0, \tag{47}$$

the variables $I_1, I_2$ stay permanently confined to close vicinities of their initial values, provided the parameter $\varepsilon$ is sufficiently small (Arnold et al. 2006).

Using the relations given in Section 2, we obtain

$$D(I_1, I_2) = -2 K_2(I_2) \cdot n_g^2(K_0(I_2), 0) + O(I_1^2)$$

under the condition $I_2 \neq I_2^d$. In the case $I_2 = I_2^d$

$$D(I_1, I_2^d) = -12 I_1^2 K_4(I_2^d) \cdot n_g^2(K_0(I_2^d), 0) + O(I_1^4).$$

Thus, for sufficiently small absolute values of the variable $I_1$, the KL Hamiltonian satisfies the condition (47). Thence, in the solutions to the studied system, the variable $I_1$ preserves its sign (and, consequently, no flips occur) if $\varepsilon < \varepsilon^*$, where $\varepsilon^*$ is a positive constant whose value depends on the values of $I_1(0)$ and $I_2(0)$.